\documentclass[prl,twocolumn,aps,amsmath,amssymb]{revtex4-1}

\usepackage{graphicx}
\usepackage{times}
\usepackage{textcomp}
\usepackage{epstopdf}
\usepackage{float}
\usepackage{bm,mathrsfs}
\usepackage[dvips]{epsfig}
\usepackage{float}
\usepackage{amsmath}
\usepackage{amssymb}
\usepackage{amsfonts}
\usepackage{enumerate}
\usepackage{hhline}
\usepackage{threeparttable}
\usepackage{multirow}
\usepackage{supertabular}
\usepackage{pslatex}
\usepackage{tabularx}
\usepackage{color}
\usepackage{graphicx}
\usepackage{dcolumn}
\usepackage{bm}

\begin{document}

\title{Quantum-limited measurement of spin qubits via curvature coupling to a cavity}

\author{Rusko Ruskov and Charles Tahan}

\affiliation{Laboratory for Physical Sciences, 8050 Greenmead Dr., College Park, MD 20740}

\email{charlie@tahan.com, ruskovr@lps.umd.edu}

\date{\today}

\begin{abstract}
We investigate coupling an encoded spin qubit to a microwave resonator
via qubit energy level curvature versus gate voltage.
This approach enables quantum non-demolition readout with
strength of tens to hundred MHz
all while the qubit stays at its full sweet-spot to charge noise,
with zero dipole moment.
A ``dispersive-like'' spin readout approach similar to circuit-QED
but avoiding the Purcell effect is proposed.
With the addition of gate voltage modulation, selective longitudinal readout
and n-qubit entanglement-by-measurement are possible.
\end{abstract}

\maketitle

{\it Introduction.-}
Quantum measurement of
semiconductor spin qubits, e.g. in quantum dots (QD),
is usually associated with the
spin-to-charge conversion technique, where spin states are mapped to  
auxiliary charge states,
of the system, and sensing the charge
is via electron transport
in a (nearby) quantum point contact or a single electron transistor
\cite{Elzerman2004N,Petta2005S,Veldhorst2015N}.
This method suffers, however,
from having to move the qubit away from its operating point,
from low sensing efficiency,
and/or susceptibility to charge  and $1/f$ noise
of the QD qubit and detector.
Thus, a readout approach
using transport of microwave (MW) photons, coupled to the spin
via a common superconducting (SC) resonator
\cite{Petersson2012N,FreyWallraff2012PRL}
and utilizing noiseless MW amplification is desirable,
as it has been proven suitable to reach a quantum-limited measurement regime
for superconducting (SC) qubits
\cite{Blais2004PRA,KorotkovSiddiqi2012N,RisteDiCarlo2013N,KorotkovSiddiqi2014PRL}.
The key
is then to establish a spin coupling to resonator  leading to
spin-dependent photon scattering.

In the standard approach, a transverse coupling
via the qubit electric dipole moment (e.d.m.), $\bm{d} \neq 0$ is used
(as in  SC transmon\cite{Blais2004PRA}),
leading to a Jaynes-Cummings interaction:
${\cal H}_{\rm tr} \simeq \hbar g_{\perp} (\sigma_{-}\hat{a}^{\dagger} + \sigma_{+}\hat{a} )$.
In the  dispersive limit,
the qubit-resonator detuning  $\Delta$ is large and  the resulting
coupling (to leading order in  $g_{\perp}/\Delta$),
${\cal H}_{\rm tr} \simeq \frac{\hbar g_{\perp}^2}{\Delta}\, \hat{a}^{\dagger} \hat{a} \sigma_z$,
commutes with the qubit Hamiltonian  ${\cal H}_q$.
Recent studies of the resonant exchange (RX) qubit \cite{Taylor2013PRL},
based on a 3-electron triple quantum dot (TQD),
have offered strong spin-cavity coupling,
at a partial sweet spot
to gate detuning fluctuations (see also Ref.\onlinecite{RussBurkard2015PRB}).

This approach, however, has several drawbacks for electron spin QD qubits:
first, for a finite
e.d.m.\ the spin qubit is more susceptible to charge noise
\cite{HuDasSarma2006PRL,Taylor2013PRL,RussBurkard2016PRB,Friesen2016preprint}.
Secondly, in higher orders of  $g_{\perp}/\Delta$, the transverse interaction
no longer commutes  with ${\cal H}_q$, dressing the qubit-resonator states
and
leading to enhanced qubit relaxation (Purcell effect) and
dressed dephasing \cite{Gambetta2009PRA,GambettaKorotkov2014PRB}
even if the resonator is
coherently populated.
These effects may increase
for a spin qubit (relative to a transmon), since the former usually possesses
a small dipole moment, and a trade off between charge noise (less e.d.m.)  and
a larger resonator photon flux (stronger measurement)
may not exist.

\begin{figure} [t!] 
    \centering
     \includegraphics[width=0.65\textwidth]{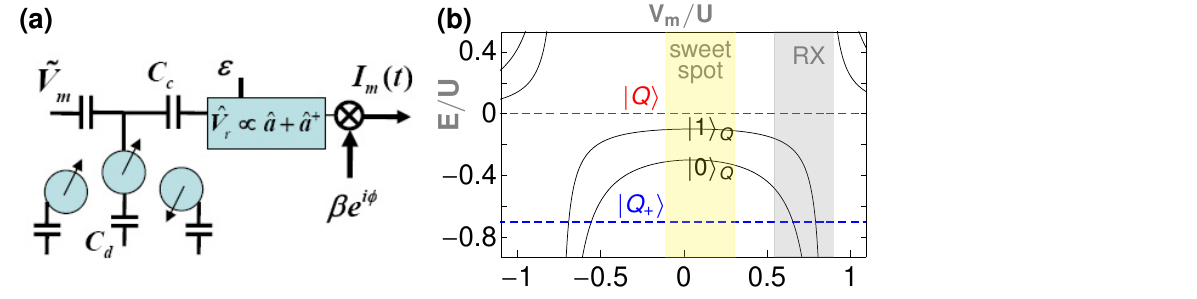}
        \caption{
        (a) A TQD
        always-on, exchange-only encoded qubit coupled to a microwave resonator, $\omega_r$,
        through the middle dot gate voltage, $V_m$.
        Readout is accomplished either
        1) via a ``dispersive-like" interaction and driving the resonator ($\epsilon$) in a manner similar to circuit-QED;
        or
        2) via modulating $V_m$ near the resonator frequency, creating a longitudinal interaction.
        Photons put into the resonator are scattered off the (far off-resonant) spin-qubit and measured, e.g.,
        via a homodyne signal, $I_m(t)$.
        (b) Qubit spin levels $E_{0,1}$ vs. gate voltage $V_m$ stay at a
        full-sweet-spot (at gate detunings $\varepsilon_v^0$, $V_m^0$, in yellow) during readout.}
        \label{fig:1}
\end{figure}

In this Letter we propose, alternatively,
a spin-to-SC-resonator  coupling,
utilizing the TQD qubit energy curvature,
$\partial^2 E_q(V_G)/\partial V_G^2$,
with a gate voltage (Fig.~1a).
We find that
a static, dispersive-like coupling
${\cal H}_{\lambda} = \hbar \lambda\omega_r \left( \hat{a}^{+}\hat{a} + \frac{1}{2} \right) \sigma_z$,
occurs due to the quantized resonator voltage.  
A dynamical longitudinal coupling:
${\cal H}_{\parallel} = \hbar \left[g_{\parallel}(t) \sigma_z + g_0(t) \hat{I} \right] (\hat{a} + \hat{a}^{+})$
also appears if an additional external voltage modulation $\tilde{V}_m(t)$
is applied  near the resonator frequency,
generating a spin-dependent ``force'' exerted on the resonator.
Similar longitudinal
coupling was explored in ion trap quantum gates \cite{Molmer1999PRL,Milburn2000F,Leibried2003N,HaljanMonroe2005PRL},
and was recently proposed by Kerman \cite{Kerman2013} and others \cite{Billangeon2015PRB,Didier2015PRL,RicherDiVincenzo2016PRB}
for SC qubits.
The energy curvature
can be appreciably large
in a regime where  e.d.m.\ is zero (Fig.~1b),
and  the charge noise to the qubit is minimized (full-sweet-spot),
previously referred to as the AEON qubit  regime \cite{AEON2016}.
The curvature interactions commute with the qubit Hamiltonian,
avoiding Purcell effect.
Importantly, we show here that quantum measurements
can be performed while each qubit
is residing at  its full sweet spot,
with a measurement time of the order of tens of ns.

{\it Dispersive-like and longitudinal curvature couplings.-}
We consider a TQD 3e-qubit coupled to a resonator via
a voltage variation on the middle dot,
$V_G = V_m^{0} + \tilde{V}_m(t) + \hat{V}_r$,
with $V_m^{0}$
at the full-sweet-spot;
$\hat{V}_r \propto \sqrt{\omega_r} \left( \hat{a}^{+} + \hat{a} \right)$
is the resonator quantized voltage, and $\tilde{V}_m(t)$ is
an external 
modulation, Fig.~1.
The Hamiltonian (including resonator  driving and environment) is:
\begin{equation}
{\cal H}_{\rm tot} = {\cal H}_{0} +
{\cal H}_{\lambda} + {\cal H}_{\parallel} +
  \varepsilon \left(\hat{a}^{+} e^{-i \omega_d t} + h.c. \right) +
{\cal H}_{\rm env} ,
\label{total_Hamiltonian}
\end{equation}
where ${\cal H}_{0} = \hbar \omega_r \hat{a}^{+}\hat{a} + {\cal H}_q $,
$\hat{a}$ is
a mode annihilation operator, and
the couplings, ${\cal H}_{\lambda}$, ${\cal H}_{\parallel}$,
derive from the qubit energy:
$
{\cal H}_q = E_g |g\rangle\langle g | + E_e |e\rangle\langle e |
\equiv
G_q(V_G) \hat{I} + \frac{E_q(V_G)}{2} \tilde{\sigma}_z
$,
expanded to second order.
With no gate modulation,
a static dispersive-like interaction ${\cal H}_{\lambda}\propto \hat{V}^2_r \sigma_z$
(neglecting contra-rotating terms $\sim \hat{a}^2,\hat{a}^{+2}$)
leads to a spin-dependent  
detuning, $\delta\omega_s=\pm \lambda\omega_r$:
\begin{equation}
\lambda \equiv \frac{\delta\omega}{\omega_r} =
\frac{\hbar \omega_r}{2} \left(\frac{\eta}{\hbar}\right)^2 \, \frac{\partial^2 E_q(V^0_m)}{e^2 \partial V_G^2} .
\label{delta-omega}
\end{equation}
The ratio
$\frac{\eta}{\hbar} \equiv \alpha_c\, \sqrt{\frac{Z_r}{\hbar/e^2}}$
includes the  middle dot coupling capacitances
to the resonator and the ground (Fig.~1a)
via  $\alpha_c \equiv \frac{C_c}{C_c + C_d}$,
and a ratio of the resonator impedance, $Z_r \simeq \omega_r L_r \gg R_r$,
to the resistance quantum.
Here, the zero-point fluctuation (for a resonator circuit, $mass \leftrightarrow L_r$)
is $\Delta x_0 = e \sqrt{\frac{\hbar/e^2}{2 \omega_r L_r}}$,
and a ratio of  $\frac{\eta}{\hbar} \gtrsim 1$  is possible for
high kinetic inductance ($L_r \gtrsim 100\, {\rm nH}$)  resonators, 
reached
in SC wires with disorder\cite{Samkharadze2016}.

By switching on a voltage modulation $\tilde{V}_m(t)$,
a term $\propto \tilde{V}_m\, \hat{V}_r$  gives 
  the  longitudinal Hamiltonian:
$
{\cal H}_{\parallel} = \hbar \left[g_{\parallel} \sigma_z + g_0 \hat{I} \right] (\hat{a} + \hat{a}^{+})
$
with couplings modulated in time:
$g_{\parallel,0} = g_{\parallel,0}^{\rm st} + \tilde{g}_{\parallel,0}(V^0_m)\, \cos ( \omega_m t + \varphi_m )$
at a frequency $\omega_m \sim \omega_r$
(a static coupling, $g_{\parallel,0}^{\rm st} \propto \frac{\partial E_q(V^0_m)}{\partial V_G}$,
is zeroed at sweet spot).
In a frame rotating with $\omega_m$,
the longitudinal couplings read:
\begin{equation}
\frac{\{ \tilde{g}_{\parallel}, \tilde{g}_{0} \} }{\omega_r}
= \frac{\eta}{e \hbar} \left\{ \frac{1}{2} \frac{\partial^2 E_q(V^0_m)}{\partial V_G^2},
 \frac{\partial^2 G_q(V^0_m)}{\partial V_G^2} \right\} \tilde{V}_m
\label{dynamic_coupling}
\end{equation}

{\it  TQD couplings estimations.-}
For a triple QD (TQD) 3-electron qubit, we are seeking
a configuration  where, ideally, the electric dipole moment is zero,
avoiding  both  spurious transverse coupling,  static longitudinal coupling,
and  also minimizing  susceptibility to charge noise.
In a recently established full-sweet-spot parameter regime \cite{AEON2016}
the relevant qubit states
are made of the two bare $(111)$-states
with spin projection $S_z=1/2$:
$|1\rangle = |S\rangle_{13} |\uparrow\rangle_2 =
\frac{1}{2}\left[|\uparrow_1 \uparrow_2 \downarrow_3 \rangle - |\downarrow_1 \uparrow_2 \uparrow_3  \rangle \right]$,
and
$|2\rangle = -\frac{1}{\sqrt{6}}
\left[|\uparrow_1 \uparrow_2 \downarrow_3 \rangle + |\downarrow_1 \uparrow_2 \uparrow_3  \rangle -
2 |\uparrow_1 \downarrow_2 \uparrow_3  \rangle \right]$
with small admixture of the other charge configurations, like $(201)$, $(102)$, etc., Fig.~1b.
Unlike the RX-regime \cite{Taylor2013PRL,Medford2013PRL},
the Coulomb energy cost, $\tilde{U}_i$, for a double occupation of the i-th dot
is large
compared to the interdot tunnelings: $\tilde{U}_i  \gg t_l, t_r$.
Coupling the resonator through the middle dot, Fig.~1a,
allows the sweet-spot   
to remain largely intact
(compare with Refs.~\cite{ReedHunter2016PRL,MartinsKuemethMarcus2016PRL}).

Introducing the  
independent  energy
detunings,
$\epsilon_v \equiv (V_3 - V_1)/2$,
and $V_m \equiv (V_3 + V_1)/2 - V_2$ ($V_i$ is the gate voltage applied to the i-th dot)
the effective qubit Hamiltonian is
recast to the form \cite{Taylor2013PRL,AEON2016}:
${\cal H}_{\rm q,eff} =
- J(\epsilon_v,V_m) \, \hat{I} - \frac{E_q(\epsilon_v,V_m)}{2} \tilde{\sigma}_z
$,
where the exchange energies
$J_{l,r}$, $J\equiv (J_l + J_r)/2$,   determine
the qubit splitting $E_q = \sqrt{J_l^2 + J_r^2 -J_l J_r}$.
At the sweet spot ($\frac{\partial E_q}{\partial \epsilon_v} = \frac{\partial E_q}{\partial V_m} =0$):
$J_{l,r}^0 \equiv J_{l,r}(\epsilon_v^0,V_m^0) = \frac{8 t_{l,r}^2}{a_{l,r} }$,
with $a_l = \tilde{U}_1 + \tilde{U}_2^{'}$, $a_r = \tilde{U}_2 + \tilde{U}_3$.
For coupling estimations,
we consider:
$a_{l,r}=a$, $t_{l,r}=t$,
resulting in    
$\frac{\partial^2 E_q^{0}}{\partial V_m^2} = 64 t^2/ a^3$.
For typical Si QD charging energies
(see Refs. \cite{Maune2012N,Medford2013PRL,ReedHunter2016PRL})
$U_1 = U_3 = U = 0.5\, {\rm meV}$, and $V_{12} = V_{23} = 0.1 U$, $V_{13}=0.05 U$,
tunneling $t=20 (40)\, \mu{\rm eV}$,
and  modulation amplitude $\tilde{V}_m = 0.1\, {\rm meV}$
one obtains
a resonator frequency shift $\delta\omega \simeq 10.3\, {\rm MHz}$
($\delta\omega = 41.4\, {\rm MHz}$),
a longitudinal
coupling $\tilde{g}_{\parallel} \simeq 25\, {\rm MHz}$
($\tilde{g}_{\parallel} = 100\, {\rm MHz}$),
and a qubit splitting $E_q = J \simeq 1\, {\rm GHz}$ ($3.9\, {\rm GHz}$),
well off resonance  
with $\omega_r/2\pi = 10\, {\rm GHz}$.
Since the scalings,
$\delta\omega_s \propto \eta^2 \omega_r^2\, t_{l,r}^2/U^3$ and $\tilde{g}_{\|} \propto \eta \omega_r\, t_{l,r}^2/U^3$
($\eta\propto\sqrt{\omega_r}$), a large range of parameters can be explored
e.g., for slightly larger dots, $U\approx 0.4\, {\rm meV}$, the couplings increase twice.
Higher
curvature corrections to $\delta\omega$, $\tilde{g}_{\parallel}$
are due to
$\partial^4 E_q^{0}/\partial V_m^4$,
and reach 5-15\%.
There appear also small non-linearities of the form,
$\zeta_{0}(\hat{a}^{+}\hat{a})^2$, $\zeta_{\|}(\hat{a}^{+}\hat{a})^2 \sigma_z$,
$\zeta_{\hat{n}}(\hat{a}^{+}\hat{n} + \hat{n}\hat{a})^2 \sigma_z$,
that are Purcell free as well.

{\it Qubit dephasing via the resonator relaxation-}
The qubit plus a SC  resonator system
is described via a Caldeira-Leggett master equation \cite{CaldeiraCerdeira1989}
(plus  qubit relaxation  and dephasing):
\begin{equation}
\frac{d\rho}{dt} = -i \left[{\cal H}_{\rm tot},\rho\right]
-i\frac{\kappa}{2\hbar} \left[\hat{x} \left\{\hat{p}, \rho \right\} \right]
-\frac{K_d}{\hbar^2}\, \left[\hat{x} \left[\hat{x}, \rho \right] \right]
\label{damping_diffusion}
\end{equation}
where $\hat{x}\equiv \Delta x_0 (\hat{a} + \hat{a}^{+})$, $\hat{p}\equiv -i \Delta p_0 (\hat{a} - \hat{a}^{+})$ are
the ``position'' and ``momentum'' operators,
and $K_d = \frac{L_r \hbar \omega_r \kappa}{2}\, \coth{\frac{\hbar \omega_r}{2 k_B T_r}}$
is the temperature dependent diffusion.
For the qubit readout
one considers
a zero resonator temperature $T_r=0$, where
under the evolution of Eq.~(\ref{damping_diffusion}),
a coherent state remains coherent (pure) state.
This
is also preserved by continuous measurements of the resonator,
(see Ref.~\cite{RuskovSchwabKorotkov2005} and references therein).

The qubit-resonator density matrix can be expanded in the complete set
of  qubit operators
$|s\rangle \langle s'|$  ($s=\pm$ for a single qubit):
$\rho = \sum_{s,s'} \hat{\rho}_{s,s'} |s\rangle \langle s'|$
where the partial   
matrices $\hat{\rho}_{s,s'}$  act only on the resonator \cite{WisemanMilburn-book2010}.
%
In this case it is sufficient to solve for the
partial density matrices, $\hat{\rho}_{s,s'}$,
using positive $P^{(+)}$-representation \cite{GardinerZoller-book2000}.
By making a Gaussian (coherent state) anzatz
$\hat{\rho}_{s,s'}(t)=\rho^q_{ss'}(t)\, |\alpha_s(t) \rangle \langle \alpha_{s'}(t) |$
for the partial density matrices \cite{Gambetta2008PRA}
one obtains the spin-dependent resonator oscillations
under  driving and modulation
(in rotating wave approximation) 
$
\dot{\alpha}_{\pm}(t) = -i \{\left[ (\tilde{\omega}_r \pm \delta\omega) -i \frac{\kappa}{2}\right] \alpha_{\pm} + \varepsilon
 + \frac{ e^{-i \varphi_m} }{2} \left[ \pm \tilde{g}_{\|} + \tilde{g}_0 \right] \}
$,
where
$\tilde{\omega}_r = \omega_r - \omega_m$, $\delta\omega \equiv \lambda\omega_r$.
For the non-diagonal qubit density matrix one gets the solution:
\begin{eqnarray}
&&  \rho^q_{+-}(t) = \rho^q_{+-}(0)\, e^{- i (\omega_q - 2 \delta\omega - i \gamma_2) t}\,
e^{- i 2 \delta\omega \int_0^t dt' \alpha_{+}(t')\, \alpha_{-}^*(t')}
\nonumber\\
&& \qquad\quad  \times
e^{- i \tilde{g}_{\|} \int_0^t dt'
\left[ \alpha_{+}(t')\, e^{i \varphi_m}  +  \alpha_{-}^*(t')\, e^{-i \varphi_m} \right] } ,
\label{a_eg-solution}
\end{eqnarray}
where
$\gamma_2\equiv 1/T_2$
is the qubit internal dephasing,
and
the last two terms can be written in the long-time limit ($t \gg 1/\kappa$)
as a resonator-induced dephasing, $e^{- \Gamma_{\rm qb,res} t}$,
with $\Gamma_{\rm qb,res} \equiv \Gamma_{\lambda} + \Gamma_{\parallel}$.
With the stationary solutions,
$
\alpha_{\pm}^{\rm st} = - \frac{\varepsilon + \frac{e^{-i \varphi_m}}{2} (\pm \tilde{g}_{\|} + \tilde{\eta}_0 ) }
{(\tilde{\omega}_r \pm \delta\omega) -i \frac{\kappa}{2}}
$, 
one obtains (at resonance $\tilde{\omega}_r = 0$ and $\varphi_m =0$) 
the dispersive-like and longitudinal contributions to $\Gamma_{\rm qb,res}$:
\begin{eqnarray}
&&  \Gamma_{\lambda} = - 2 \delta\omega\, {\rm Im} \left[\alpha_{+}^{\rm st} \alpha_{-}^{\rm st*} \right]
\nonumber\\
&& \qquad { } = \frac{(2\delta\omega)^2\, (\kappa/2)}{\left[ \delta\omega^2 + \kappa^2/4 \right]^2}
\left[ \left(\varepsilon + \tilde{g}_0 /2 \right)^2 - \left( \tilde{g}_{\|} /2 \right)^2\right]  \qquad
\label{gamma_qb_lambda}
\\
&& \Gamma_{\parallel} = - \tilde{g}_{\|} \,
{\rm Im} \left[\alpha_{+}^{\rm st}  +  \alpha_{-}^{\rm st*} \right]
= \frac{\tilde{g}_{\|}^2\, (\kappa/2)}{ \delta\omega^2 + \kappa^2/4 } . 
\label{gamma_qb_parallel}
\end{eqnarray}
One also obtains
qubit frequency (ac Stark) shifts:
$
\delta\omega_{\lambda}
= - 2 \delta\omega\, {\rm Re} \left[\alpha_{+}^{\rm st} \alpha_{-}^{\rm st*} \right]
= \frac{ 2\delta\omega \left[ \delta\omega^2 - \kappa^2/4 \right]}{\left[ \delta\omega^2 + \kappa^2/4 \right]^2}
\left[ \left(\varepsilon + \tilde{g}_0/2 \right)^2 - \left( \tilde{g}_{\|}/2 \right)^2\right]
$,
$
\delta\omega_{\|}
= -\tilde{g}_{\|} {\rm Re} \left[\alpha_{+}^{\rm st}  +  \alpha_{-}^{\rm st*} \right]
= \frac{\tilde{g}_{\|}^2\, \delta\omega}{ \delta\omega^2 + \kappa^2/4 }
$.

{\it  Longitudinal readout-}
The rate $\Gamma_{\rm qb,res}$ can be interpreted as
the maximal measurement rate of a qubit.
Indeed, performing a homodyne measurement of the resonator ,
the measurement signal is \cite{WisemanMilburn-book2010}:
$
I_{\rm m}(t) = \beta \left[ \kappa \langle \hat{X}_{\phi} \rangle + \sqrt{\kappa} \xi(t)\right]
$,
where $\hat{X}_{\phi} = \hat{a}\, e^{-i\phi} + \hat{a}^{+}\, e^{i\phi}$ is the measured
resonator quadrature,
$\langle \ldots \rangle \equiv \mbox{\rm Tr} [ \rho_m \ldots ]$ is the quantum average
($\rho_m$ is the conditioned system density matrix),
$\beta$ and $\phi$  are respectively the strength and  phase   of the local oscillator (Fig.~1a),
and $\xi(t)$  is the detector noise.
In the
``bad cavity limit'' (when $\kappa \gg \Gamma_{\rm qb,res}$),
the photon leakage out of the resonator is much faster than the qubit internal evolution,
implying that homodyne measurement of the resonator field is a qubit measurement.
In the stationary regime, for fixed qubit states $|\pm\rangle$,
the average currents read
\begin{equation}
I_{\pm} = \beta \kappa \left[ \alpha_{\pm}^{\rm st}\,e^{-i\phi} + \alpha_{\pm}^{\rm st*}\,e^{i\phi} \right],
\label{average currents}
\end{equation}
and the current signal can be expressed via the qubit  (conditioned) density matrix $\rho^q_m(t)$,
with $\Delta I \equiv I_{+} - I_{-}$:
\begin{equation}
I_m(t) = \frac{\Delta I}{2} {\rm Tr} [\rho^q_m(t) \sigma_z] + \beta \sqrt{\kappa} \xi(t) .
\label{qubit_homodyne_signal}
\end{equation}
The (single sided) current spectral density, $S_0$, is related to
the photon shot noise $I_0(t) \simeq  \beta \sqrt{\kappa} \xi(t)$ via:
$\langle I_0(t)\, I_0(t')\rangle_{\xi} = \beta^2 \kappa \delta(t-t') \equiv \frac{S_0}{2}\, \delta(t-t')$.
For each of the  
qubit states $|+\rangle$, $|-\rangle$, the  random
finite time average  $\overline{I}(t) = \frac{1}{t} \int_0^t I_{\rm m}(t') dt'$
is Gaussian distributed with the averages $I_{\pm}$ and variance
$\frac{S_0}{2 t}$
[assuming weak
response, $\Delta I \ll \beta\kappa$].
Then, a measurement time $\tau_{\rm meas} \equiv 1/\Gamma_{\rm m}$  can be introduced,
as the time needed to distinguish between the two state-dependent
distributions \cite{KorotkovPRB2001}:
$\tau_{\rm meas} = \frac{4 S_0}{\left( I_{+} - I_{-} \right)^2}$.
The detector response takes the form
$\Delta I = 2 \beta\kappa |\alpha_{+}^{\rm st} - \alpha_{-}^{\rm st} |\, \cos{\Delta\phi}$,
and is maximized
by choosing the measured quadrature $\hat{X}_\phi$, so that
$\Delta\phi \equiv \phi - \phi_{\rm dif}=0$,
where $\phi_{\rm dif}={\rm arg}[\alpha_{+}^{\rm st} - \alpha_{-}^{\rm st}]$ is the phase of the
difference for the two resonator fields.
Thus, the maximal measurement rate is:
\begin{equation}
\Gamma_{\rm m}^{\rm max} = \frac{\Delta I_{\rm max}^2}{4 S_0} = |\alpha_{+}^{\rm st} - \alpha_{-}^{\rm st} |^2\, \frac{\kappa}{2} ,
\label{measurement_rate_max}
\end{equation}
which is equal to the dephasing rate,
$\Gamma_{\rm qb,res} = \Gamma_{\lambda} + \Gamma_{\parallel}$,
of Eqs.~(\ref{gamma_qb_lambda}),(\ref{gamma_qb_parallel}).
For quadrature measurement with $\phi \neq \phi_{\rm dif}$ one gets
a rate
$\Gamma_{\rm m} = \Gamma_{\rm m}^{\rm max} \cos^2{(\phi - \phi_{\rm dif})}$
[see also Ref.~\cite{Gambetta2008PRA}].

The qubit density matrix at time $t$, given the  measurement
record   $\overline{I}(t)$,
will be updated  
via a quantum Bayesian rule \cite{KorotkovPRB2001}: 
\begin{equation}
\rho^q_{\rm m}(t,\overline{I}) =
\frac{\hat{U}_{\overline{I},z}\, \hat{M}_{\overline{I},z} \rho^q(0) \hat{M}_{\overline{I},z}^{+} \hat{U}_{\overline{I},z}^{+} }
{P[\overline{I}(t)]} ,
\label{Bayesian_rule}
\end{equation}
where
$P[\overline{I}(t)] \equiv {\rm Tr} [ \rho^q(0)\, \hat{M}_{\overline{I},z}^{+}\, \hat{M}_{\overline{I},z} ]$
is the total probability of the ``event'' $\overline{I}(t)$,
the measurement operators are
$
\hat{M}_{\overline{I},z} = ( \frac{t}{\pi S_0} )^{\frac{1}{4}}\,  
e^{ - [\overline{I}(t) - \frac{\Delta I_{\rm max}}{2}\cos\Delta\phi\, \hat{\sigma}_z]^2 \,\frac{t}{2 S_0} }
$,
and  an additional
 unitary backaction is induced  by the homodyne
measurement \cite{WisemanMilburn-book2010,Gambetta2008PRA,Korotkov2016PRA94}:
$\hat{U}_{\overline{I},z} =
e^{i  \overline{I}(t)\, \frac{\Delta I_{\rm max}}{2} \sin\Delta\phi\, \hat{\sigma}_z\, \frac{t}{S_0}} e^{\!-i \hat{\varphi}_1(t)}$.
Measuring the quadrature with $\phi = \phi_{\rm dif}$ eliminates this  backaction
(except the deterministic phases $\hat{\varphi}_1(t)$, see below)
and leads
to maximum information inferred from the qubit.
The
measurement operators, $\hat{M}_{\overline{I},z}$
are derived from two requirements:
(i) the qubit diagonal density matrix elements $\rho^{\rm m}_{++}(t)$,
$\rho^{\rm m}_{--}(t)$, are updated according to a  classical Bayesian rule
with likelihood  probabilities
$P_{\pm}(\overline{I}) =
( \frac{t}{\pi S_0} )^{\frac{1}{2}}   
e^{ - (\overline{I}(t) \mp \frac{\Delta I_{\rm max}}{2}\cos\Delta\phi)^2 \, \frac{t}{S_0} }$,
obeying a quantum-classical correspondence \cite{KorotkovPRB2001};
(ii)
the evolution of the non-diagonal element obeys the rule:
$\rho^{\rm m}_{+-}(t) = \rho^{\rm m}_{+-}(0)\sqrt{\rho^{\rm m}_{++}(t)\rho^{\rm m}_{--}(t)/\rho^{\rm m}_{++}(0)\rho^{\rm m}_{--}(0)}$,
which follows from the saturation of the inequality $|\rho_{ij}| \leq \sqrt{\rho_{ii}\rho_{jj}}$ averaged over
all possible records $\overline{I}(t)$ \cite{KorotkovPRB2001}.
By differentiating Eq.~(\ref{Bayesian_rule}) a stochastic evolution equation can be obtained.

{\it  $n$-qubit readout-}
Consider simultaneous measurement of $n$ qubits coupled to a  resonator.
The basis spin states are of the form $|s \rangle = |\uparrow\uparrow \dots \downarrow \rangle$.
The $n$-qubit couplings $G_{\parallel, s}$ and detunings $\delta\omega_s$ appear to be:
$
G_{\parallel, s} =
 \langle s|\, \frac{ e^{-i \varphi_m} }{2}  \sum_j
\left[ \tilde{g}^{(j)}_{0} + \tilde{g}^{(j)}_{\parallel} \sigma_z^{(j)}  \right] \, | s\rangle
$,
and  $\delta \omega_s =  \langle s| \sum_j \delta\omega_j \sigma_z^{(j)} | s\rangle$.
With the solution of
$
\dot{\alpha}_s(t) = -i \{\left[ (\tilde{\omega}_r + \delta\omega_s) -i \frac{\kappa}{2}\right] \alpha_s + \varepsilon
 + G_{\parallel, s} \}
$
one gets the average currents
$I_{\rm s}(\phi) = \beta \kappa \left[ \alpha_s^{\rm st}\,e^{-i\phi} + \alpha_s^{\rm st*}\,e^{i\phi} \right]$
and the accumulated phases $\varphi_s(t)$,
$\dot{\varphi}_s = {\rm Re} [(\varepsilon + G_{\parallel, s})^* \alpha_s(t)]$
for each state $|s\rangle$.
The $n$-qubit measurement operator   
reads
($\hat{\Omega}_{\overline{I},\hat{c}_n} \equiv  \hat{M}_{\overline{I},\hat{c}_n} \cdot \hat{U}_{\overline{I},\hat{c}_n}$):
\begin{equation}
\hat{\Omega}_{\overline{I},\hat{c}_n(\phi)} =
{\textstyle ( \frac{t}{\pi S_0} )^{\frac{1}{4}} }   
e^{ - [\overline{I}(t) - \hat{c}_n(\phi)]^2 \, \frac{t}{2 S_0} } \cdot
e^{ i  \overline{I}(t)\,  \hat{c}_n(\phi+\frac{\pi}{2})\frac{t}{S_0}}  e^{\!\! -i \hat{\varphi}_n(t)}
\label{n-qubits-meas-operator}
\end{equation}
with $\hat{c}_n(\phi) = {\rm diag}\{I_1{\scriptstyle (\phi)}, \ldots\!, I_{2^n}{\scriptstyle (\phi)} \}$,
$\hat{\varphi}_n(t) = {\rm diag}\{\varphi_1, \ldots\!, \varphi_{2^n} \}$
being diagonal operators.
In Eq.~(\ref{n-qubits-meas-operator}) the ``pure'' measurement operator $\hat{M}_{\hat{c}_n}$ is in general form
and
is equivalent to a quantum Bayesian update of the $n$-qubit density matrix,
while $\hat{U}_{\hat{c}_n}$
is a unitary backaction
derived using a ``history tails'' approach \cite{Korotkov2016PRA94}.
By  averaging over all possible realizations of $\overline{I}(t)$ one obtains
the ensemble evolution:
$\dot{\rho}^q_{ss'} = ( -i \omega_{ss'} - \Gamma_{ss'} ) \rho^q_{ss'}$, where
$\Gamma_{ss'} = \frac{\Delta I_{{\rm max};ss'}}{4 S_0} = |\alpha_s^{\rm st} - \alpha_{s'}^{\rm st}|^2\frac{\kappa}{2}$
are the partial dephasing rates,  analogous to Eq.~(\ref{measurement_rate_max}),
and
$\omega_{ss'} = (\delta\omega_s - \delta\omega_{s'})\, {\rm Re} \left[\alpha_{s}^{\rm st} \alpha_{s'}^{\rm st*} \right]
+
{\rm Re} \left[ (G_{\parallel, s} - G_{\parallel, s'})^* ( \alpha_{s}^{\rm st} + \alpha_{s'}^{\rm st*} ) \right]$
are the
ac-Stark shifts, similar to the single-qubit case.

By performing a joint measurement of qubits by a single resonator one can entangle them without a direct
qubit interaction \cite{RuskovKorotkov2003PRB67}. In the simplest case of two qubits this is achieved
provided the measurement cannot distinguish certain two-qubit subspaces
\cite{RuskovKorotkov2003PRB67,MaoAverinRuskovKorotkov2004PRL,RuskovKorotkovMizel2006PRB72,RisteDiCarlo2013N,KorotkovSiddiqi2014PRL}.
Assuming equal couplings, $\lambda_1=\lambda_2$, $\tilde{g}_{\|,1}=\tilde{g}_{\|,2}$
(i.e. symmetric measurement),
one gets $I_{\uparrow\downarrow} = I_{\downarrow\uparrow}$, a necessary condition
for two-qubit entanglement by a joint measurement \cite{RuskovKorotkov2003PRB67}.
In the limit, $\kappa \gg \delta\omega,\tilde{g}_{\|}$ the current differences are
equal:
$I_{\downarrow\downarrow} - I_{\uparrow\downarrow} = -(I_{\uparrow\uparrow} - I_{\uparrow\downarrow})
= 4\beta \tilde{g}_{\|} \sin \phi $, corresponding to a linear detector response
which leads to an effective measurement of the total spin $(\overrightarrow{\sigma}_1 + \overrightarrow{\sigma}_2)^2$
and probabilistic entanglement, e.g., to the spin-zero subspace,
starting from any separable initial state \cite{RuskovKorotkov2003PRB67}.
By choosing the local oscillator phase $\phi=\pi/2$ one can eliminate the additional unitary backaction,
while maximizing the qubit response.

For measurement rate estimation
one considers the ``bad cavity limit'', when $\omega_r \gg \kappa \gg \Gamma_{\rm m}$,
also  requiring $\kappa \gg \delta\omega, \tilde{g}_{\parallel}$.
With an external driving $\epsilon \approx 20 (15)\, {\rm MHz}$
(at $\omega_d=\omega_r$),
no modulation,
and for the parameters of $\omega_r/2\pi = 5 (10)\, {\rm GHz}$, $\eta/\hbar = 0.35 (1.)$,
tunneling $t_{l,r} = 40 (20)\, \mu{\rm eV}$, $U_{\rm charge}=0.32 (0.5)\, {\rm meV}$, and
loaded resonator $Q$-factor $1.25 (2.5) 10^2$,
one gets:
$E_q \approx 6(1)\, {\rm GHz}$,
$\delta\omega = 5.2 (10.3)\, {\rm MHz}$,
and $\kappa/2\pi = 40\, {\rm MHz}$, thus obtaining
$\Gamma_{\rm m}/2\pi \simeq 4.7 (7.4)\, {\rm MHz}$,
average photon number $\bar{n} \approx 0.9 (0.4)$,
and a measurement time of $\tau_m \simeq 34 (21.3)\, {\rm ns}$.
Similarly,
with modulation $\tilde{V}_m = 0.1\, {\rm meV}$
(at $\omega_m=\omega_r$) and no  driving,
for the parameters  $\omega_r/2\pi = 5 (10)\, {\rm GHz}$, $\eta/\hbar = 0.35 (1.)$,
tunneling $t_{l,r} = 20\, \mu{\rm eV}$, $U_{\rm charge}=0.4 (0.5)\, {\rm meV}$,
and $Q = 10^2$, one gets:
$E_q \approx 1\, {\rm GHz}$,
$\delta\omega = 0.6 (10.3)\, {\rm MHz}$, $\tilde{g}_{\parallel} = 8.8 (25)\, {\rm MHz}$,
and $\kappa/2\pi = 100\, {\rm MHz}$, thus obtaining
$\Gamma_{\rm m}/2\pi \simeq 3.1 (13.4)\, {\rm MHz}$,  
photon number of $\bar{n} \approx 0.3 (0.5)$,
and a measurement time of $\tau_m \simeq 50.8 (11.8)\, {\rm ns}$,
much smaller than typical $T_2$ of tens of $\mu{\rm s}$ \cite{Medford2013PRL,ReedHunter2016PRL}.

{\it  Summary.-}
The proposed dispersive-like and longitudinal curvature couplings  of a TQD spin-qubit to a SC resonator
of tens to hundred MHz can be much larger  than the
transverse dispersive coupling 
$g_{\perp}^2/\Delta$
for a similar TQD system \cite{Taylor2013PRL,RussBurkard2015PRB},
which needs a large e.d.m.
These couplings can be  comparable to
superconducting qubits \cite{Wallraff2017preprint}, allowing
fast spin readout  at tens of ns.
As opposed to
Jaynes-Cummings interaction,
curvature couplings  are Purcell free,
admitting higher photon numbers and even shorter readout times.
The curvature couplings allow
for spin measurements at a sweet-spot
\footnote{These can be used in a recently proposed
quadrupole qubit \cite{Friesen2016preprint}.},
with zero QDs'
e.d.m.\!\!
and minimized qubit  dephasing,
allowing for high readout efficiency.
With the  dispersive-like coupling  $\lambda\omega_r$,
a quantum-limited  readout of individual qubits can be performed, as in CQED.
On the other hand, in a regime where
$\lambda\omega_r \ll \tilde{g}_{\|}$,
and using the $n$-qubit measurement result Eq.~(\ref{n-qubits-meas-operator}),
it is possible to utilize designated resonator(s) that selectively  couple to a number of spin-qubits,
which could be a viable route
to generate spin entanglement within a cluster of qubits,
and
to create medium range
spin entanglement across chip,
which can be a resource in quantum computations \cite{Briegel2001PRL}.

%

\end{document}